\documentclass[jcp,aip,showpacs,amsmath,amssymb,unsortedaddress,12p]{revtex4-1}

\usepackage{natbib}
\usepackage{graphics}
\usepackage{amsfonts}
\usepackage{amssymb}
\usepackage{amsmath}

\begin{document}

\title{Ion-specific colloidal aggregation: \\ population balance equations and potential of mean force}

\author{Gerardo Odriozola}
\email{godriozo@imp.mx} 

\affiliation{Programa de Ingenier\'{\i}a
Molecular, Instituto Mexicano del Petr\'{o}leo, Eje Central
L\'{a}zaro C\'ardenas 152, 07730, M\'{e}xico, Distrito Federal,
M\'{e}xico.}

\date{\today}

\begin{abstract}
Recently reported colloidal aggregation data obtained for different monovalent salts (NaCl, NaNO$_3$, and NaSCN) and at high electrolyte concentrations are matched with the stochastic solutions of the master equation to obtain bond average lifetimes and bond formation probabilities. This was done for a cationic and an anionic system of similar particle size and absolute charge. Following the series Cl$^-$, NO$_3^-$, SCN$^-$, the parameters obtained from the fitting procedure to the kinetic data suggest: i) The existence of a potential of mean force (PMF) barrier and an increasing trend for it for both latices. ii) An increasing trend for the PMF at contact, for the cationic system, and a practically constant value for the anionic system. iii) A decreasing trend for the depth of the secondary minimum. This complex behavior is in general supported by Monte Carlo simulations, which are implemented to obtain the PMF of a pair of colloidal particles immersed in the corresponding electrolyte solution. All these findings contrast the Derjaguin, Landau, Verwey, and Overbeek theory predictions.   
\end{abstract}

\maketitle

\section{Introduction}
An extremely brief view of the well established colloidal aggregation picture 
under high electrolyte concentrations could be as follows~\cite{Hunter}: At 
sufficiently high electrolyte concentrations the repulsive electrostatic 
contribution to the Derjaguin, Landau, Verwey, and Overbeek (DLVO) potential 
turns negligible, and so, the colloid-colloid Hamaker attractive contribution 
dominates the effective interaction. Thus, all colloid-colloid (and 
cluster-cluster) collisions lead to the formation of irreversible and rigid 
bonds producing the so called diffusion limited cluster aggregation (DLCA) 
regime. The minimum electrolyte concentration needed to produce DLCA (in 
practice, to produce the maximum overall aggregation kinetics) is called 
critical coagulation concentration. Since this concentration corresponds to the 
total screening of the electrostatic contribution of the DLVO potential, larger 
amounts of electrolyte do not change the obtained DLCA kinetics. 

In the above described view, valence and hydrated ionic size are considered 
(without taking into account the ion-ion short range correlations), while the 
nature of the employed electrolyte is completely disregarded. This contrasts a 
very large number of experimental observations which point out the specificity 
of some effective ion-surface interactions. Indeed, it has been clear for over a 
century the existence of systematic ion effects (widely known as Hofmeister 
effects~\cite{Collins85,Cacace97,Boinovich10}), which are strongly dependent on 
the ionic nature. These effects refer to the specificity manifested by certain 
ions on a plethora of phenomena, including surface tension at the air-water 
interface, heats of hydrations, stability and solubility of proteins, etc. A 
full and precise description of these effects must consider ion-surface, 
ion-ion, ion-water, water-surface, water-water, and direct surface-surface 
interactions~\cite{Tavares04,Manciu05,Bostrom05,Quesada-Perez10,Kalcher10,
Boinovich10}. All these contributions are not additive and so, mathematical 
treatments should not consider them independently. 

In recent work, more evidence was found pointing out the specificity of ion 
effects~\cite{Lopez-Leon10}. In this case it was shown that colloidal 
aggregation kinetics of hydrophobic colloidal particles at high monovalent 
electrolyte concentrations is extremely sensitive to the nature of the anion. 
That is, Cl$^-$ was found to produce the expected DLCA-like regime, whereas 
SCN$^-$ at the same concentration produced a steady-state cluster size 
distribution (CSD). In this paper these experimental data are matched with the 
stochastic solutions of the master equation to gain further insight into the 
process kinetics. The produced parameters, bond average lifetimes and primary 
bond formation probabilities, point out to the existence of a colloid-colloid 
potential of mean force (PMF) barrier for all employed monovalent electrolytes 
(NaCl, NaNO$_3$, and NaSCN) and systems (positive and negative colloids). 
Furthermore, also for positive and negative colloidal particles, they suggest a 
shallowing trend for the PMF well depth and an increasing trend for the PMF 
barrier following the series Cl$^-$, NO3$^-$, SCN$^-$. Thus, Monte Carlo 
simulations were implemented to see whether or not these trends can be captured. 
For this purpose, the potentials of mean force of a pair of colloidal particles 
immersed in the corresponding electrolyte solutions are calculated by including 
colloid-ion and ion-ion dispersion contributions. As shown in the results 
section, the general trends suggested by the population balance analysis agree 
with those obtained from simulations. 

The paper is structured as follows: Section~\ref{PBF} summarizes the employed 
methodology to extract bond average lifetimes and primary bond formation 
probabilities from the experimental CSDs. This section also presents the 
obtained fitted values. Section~\ref{Sim} gives details on the employed MC 
method to obtain the PMF of two colloidal particles immersed in an electrolyte 
solution, in correspondence with the experimental conditions. Sec.~\ref{Res} 
presents the MC results and links them with the obtained bond average lifetimes 
and primary bond formation probabilities. Conclusions are drawn in 
Sec.~\ref{Conc}.           

\section{Population Balance Fitting}\label{PBF}

\subsection{Theoretical background}

In order to gain physical information from the experimental time evolutions of 
the CSD, the stochastic master equation~\cite{gillespie77,thorn94,thorn95} 
corresponding to a reversible aggregation model, including both aggregation and 
fragmentation kernels~\cite{elminyawi91,pefferkorn98,Feng08}, is solved to match 
them. This master equation is the stochastic analogous to the deterministic 
population balance equations~\cite{smol16,smol17} (a detailed description of the 
model and the algorithm employed to stochastically produce the CSDs is given in 
reference~\cite{Odriozola03} section 4.3). The model behind the mathematical 
treatment assumes that two kinds of bonds, primary and secondary, can be formed. 
Primary bonds occur in an energy minimum that is very close to the particle 
surface, and then, is associated to interactions between bare particles. 
Secondary bonds occur at a certain distance from the particle surface, and thus, 
refer to situations where an energetic barrier prevents particles from 
completely approaching. These two kinds of bonds have different breakup 
probabilities and are treated separately. The energetic barrier enters as the 
probability, $P_1$, to form a primary bond given that a bond is formed (thus, 
the probability for producing a secondary bond given that a bond is formed is 
$1-P_1$). Furthermore, since the model assumes no barrier to form the secondary 
bonds, all collisions are effective, i.~e., collisions always lead to either 
primary or secondary bond formation. Therefore, the Brownian kernel can be used 
to model the aggregation kinetics of the system. This kernel is given by 
\begin{equation}\label{Bronwian}
k^{Brow}_{ij}=\frac{1}{4}k_{11}(i^{1/d_f}+j^{1/d_f})(i^{-1/d_f}+j^{-1/d_f})
\end{equation}
where $k_{11}$ is the dimmer formation rate constant, and $d_f$ is the 
clusters' fractal dimension. The fragmentation kernel $f_{ij}$ is given by 
\begin{equation}
f_{ij}=e_{ij}(1\!+\delta_{ij})(1\!-\!P_c)\left(\frac{E_1}{\tau_1(E_1\!+\!E_2)}\!+\!\frac{E_2}{\tau_2(E_1\!+\!E_2)}\right)
\end{equation}
where $E_1$ and $E_2$ are the number of primary and secondary bonds in the 
system, $\tau_1$ is the average lifetime of primary bonds, $\tau_2$ is the 
average lifetime of secondary bonds, and $\delta_{ij}$ is the Kronecker delta 
function. $e_{ij}$ is the average number of bonds that, after breaking-up, leads 
to $i-$ and $j-$size fragments. This function was approached by averaging over 
all fragmentation possibilities of a vast collection of simulated cluster 
structures and is given in reference~\cite{ruptura01}. Finally, 
$P_c=1-0.164(ij)^{-0.35}$ is the probability for two just produced clusters to 
collide and re-aggregate~\cite{europhysics01,Lattuada03,Lattuada04,Lattuada06}. 
Both kernels are used to obtain the time evolution of the CSD by stochastically 
solving the population balance equations as explained in 
reference~\cite{ruptura03}. As mentioned, $P_1$ is introduced in order to 
discern whether a primary or secondary bond is formed when a cluster-cluster 
collision occurs. The values of the parameters $k_{11}$, $P_1$, $\tau_1$, and 
$\tau_2$ result from fitting the solutions of the population balance equations 
to the experimental CSDs. However, $k_{11}$ must have a fixed value 
independently of the employed anion (since the model states that every collision 
leads to bond formation and the viscosity variations are negligible). The best 
overall fits are obtained for $k_{11}=9.0\times10^{-18}$ m$^3/$s, which is a 
value within the boundaries of the $k_{11}$ range generally accepted for 
diffusion limited cluster aggregation~\cite{sonntag87}, 
$k_{11}=(6.0\pm3.0)\times10^{-18}$ m$^3/$s. The parameter $d_f$ was fixed to the 
typical value of DLCA, $d_f=1.8$~\cite{meakin98}, for the Cl$^{-}$ ion, and to 
$d_f = 2.0$ in the other cases. $P_1$, $\tau_1$, and $\tau_2$ were considered as 
free parameters. It should be noted that, by handling these parameters, the two 
classical aggregation regimes can be reproduced: i) DLCA~\cite{lin90pcm} when 
$P_1=1$  and $\tau_1 \rightarrow \infty$, for all $\tau_2$ values; or $\tau_1 
\rightarrow \infty$, and $\tau_2 \rightarrow \infty$ for all $P_1$; or $P_1=0$ 
and $\tau_2 \rightarrow \infty$, for all $\tau_1$ and ii) Reaction limited 
cluster aggregation (RLCA)~\cite{lin90pra} when $\tau_2 = 0$ and $\tau_1 
\rightarrow \infty$, being $P_1$ the sticking probability. Therefore, this 
reversible model contains DLCA and RLCA as limiting cases. These limiting cases 
were tested to make sure the correctness of the implemented algorithm.

\subsection{Fitted curves and parameters ($P_1$, $\tau_1$, and $\tau_2$)}

\begin{figure}
\resizebox{0.5\textwidth}{!}{\includegraphics{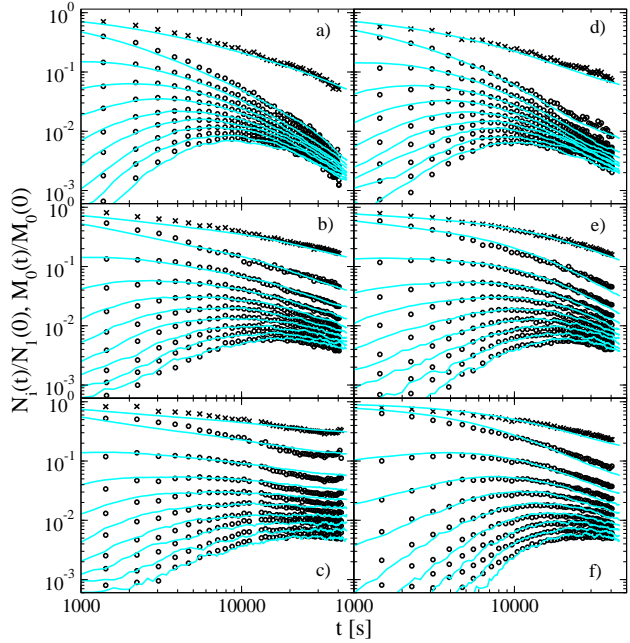}}
\caption{\label{CSD} Time evolutions of the normalized CSDs, 
$N_i(t)/N_1(t\!=\!0)$, for the aggregation induced by a) NaCl, b) NaNO$_3$, and 
c) NaSCN and positive colloidal particles. Panels d), e), and f) correspond to 
the aggregation of negative colloidal particles under the same electrolytes, 
respectively. Open circles are the experimental CSDs (from monomers up to 
clusters having nine particles) whereas crosses correspond to the normalized 
total number of clusters, $M_0(t)/M_0(t\!=\!0)$ (data taken 
from~\cite{Lopez-Leon10}). Lines are the theoretical fits.}
\end{figure}

\begin{table*}
\caption{$P_1$, $\tau_1$, and $\tau_2$ values of both latices when aggregation 
is induced by different electrolytes. Fitting errors are less than 20$\%$ in all 
cases. Mobility data, $\mu_e$, are also included (taken 
from~\cite{Lopez-Leon10}).} \label{Table1}
\begin{tabular}{||c|c|c|c|c|c|c|c|c||}
\hline
\hline
& \multicolumn{4}{c|}{Cationic} & \multicolumn{4}{c|}{Anionic} \\
\hline
& $\mu_e$ $10^8$ [m$^2/$Vs] & $\tau_1$ [s] & $\tau_2$ [s] & $P_1$ & $\mu_e$ $10^8$ [m$^2/$Vs] & $\tau_1$ [s] & $\tau_2$ [s] & $P_1$ \\
\hline
NaCl  &  $ 0.79\pm0.18$ & $>1\times10^6$ & 2000 & 0.33 & $-1.17\pm0.10$ & 200000 & 400 & 0.30 \\
NaNO3 &  $-0.03\pm0.19$ & 100000         & 600  & 0.07 & $-1.27\pm0.10$ & 200000 & 150 & 0.06 \\
NaSCN &  $-0.90\pm0.09$ & 25000          & 370  & 0.04 & $-1.37\pm0.10$ & 200000 & 100 & 0.04 \\
\hline
\hline
\end{tabular}
\end{table*}

As mentioned, the probability of forming a primary bond given a 
particle-particle collision, $P_1$, and the average lifetime of primary and 
secondary bonds, $\tau_1$, and $\tau_2$, are taken as free parameters to fit the 
experimental CSDs. The resulting curves  ($N_i(t)/N_1(t\!=\!0)$, being $N_i(t)$ 
the number of $i$-size clusters at time $t$) are plotted in panels a)-c) of 
Fig.~\ref{CSD} for the positively charged particles and in panels d)-f) of the 
same figure for the negatively charged ones, where the points refer to the 
experimental data and the lines to the stochastic solutions of the master 
equation. In this figure, panels a) and d), b) and e), and c) and f) show the 
data obtained under NaCl, NaNO$_3$, and NaSCN, respectively. Additionally, all 
panels show the normalized total number of clusters, $M_0(t)/M_0(t\!=\!0)$ 
($M_0(t)=\sum_{i=1}^{i=\inf} N_i(t)$), as crosses (experimental) and cyan lines 
(fits). The obtained values of the fitted parameters are listed in Table 1. The 
obtained agreement between experimental results and theoretical fits is good for 
all cases. 

It is common saying that three parameters are enough to fit practically any 
well behaved curve. So, the question -how much can we trust the values of the 
fitted parameters?- naturally arises. To answer it one should take into account 
that not only a single curve, but a very important part of the whole CSD is 
being fitted with the employed parameters (oligomers evolution plus the total 
number of clusters). This adds much difficulty to the fitting procedure. 
Furthermore, the parameters have physical meaning and consequently cannot take 
any value. That is, $P_1$ is restricted to $0\leq P_1 \leq1$, and $0\leq \tau_2 
\leq \tau_1$. Once that is said, it should be pointed out some limitations of 
the employed model. The construction of function $e_{ij}$ is based on loop-less 
aggregates hiving a fixed fractal dimension, $d_f\approxeq 2.0$. On the one 
hand, loop-less aggregates imply that all bond breaking events lead to cluster 
fragmentation. On the other hand, $e_{ij}$ is expected to increase with $d_f$. 
Both assumptions (loop-less aggregates and $d_f=2.0$) may not correspond to 
reality when bonds allow for the relative movement among the particles of a 
cluster (restructuring)~\cite{Tirado-Miranda99,Babu08}. In this case, the 
clusters fractal dimension raises probably reaching values over 2.0. This in 
turn enters in equation~\ref{Bronwian}, for which its solutions are luckily not 
very sensitive to this parameter~\cite{Babu08} (a larger $d_f$ produce smaller 
cross sections which practically compensates the larger diffusion coefficients, 
although the small-large aggregation turns relatively less favorable). 
Notwithstanding, the fitted parameter values surely shift when restructuring 
occurs. Finally, it should also be mentioned that the employed method of 
hydrodynamic focusing of clusters, needed for obtaining the detailed 
experimental data, probably enhances breakup. For all these reasons, it is safer 
to consider trends to be reliable only.  

When Cl$^-$ acts as the counter-ion, i.~e., for positive particles, a DLCA 
model with $k_{11} = 6.0 \times 10^{-18}$ m$^3$/s ($P_1 = 1$ and $\tau_1 > 
1\times 10^6$ s) provides a relatively good agreement with the experimental data 
(not shown). Nevertheless, the best fit is obtained for $k_{11} = 
9.0\times10^{-18}$ m$^3$/s, $P_1$ = 0.33, $\tau_1 > 1\times 10^6$, and $\tau_2 = 
2000$ s, suggesting that, even for the fastest aggregation kinetics, there is 
always a small potential barrier that avoids reaching total effectiveness in the 
collisions between particles. Something similar occurs for the negative system 
when Na$^+$ acts as the counterion (see Table~\ref{Table1}). Although 
introducing extra fitting parameters is not an absolute requisite to fit the CSD 
induced by NaCl for both systems, it becomes imperative when NO$_3^{-}$ or 
SCN$^-$ act as the counter-ions. The CSDs induced by these anions cannot be 
explained without considering the formation of reversible bonds.

For positive particles and when NO$_3^-$ is the counter-ion (Fig.~\ref{CSD} 
b)), the values of $P_1$, $\tau_1$, and $\tau_2$ importantly drop off:  $P_1 = 
0.07$, $\tau_1 =100000$ s, and $\tau_2 = 600$ s. The pronounced decrease of 
$P_1$ signals an increase in the number of secondary bonds, whose lifetimes 
become also shorter. This would translate into a higher mean force potential 
barrier and a shallower secondary minimum. This trend is confirmed by the 
analysis of the anion having the larger dispersion contribution, SCN$^-$. In 
this case the rate between secondary and primary bonds induced by SCN$^-$ 
increases with respect to NO$_3^-$, and the lifetime of the bonds decreases, 
revealing the existence of weaker bonds between particles: $P_1 = 0.04$, $\tau_1 
= 25000$ s, and $\tau_2 = 370$ s. Actually, in the regime induced by SCN$^-$, a 
balance between the number of new formed bonds and broken bonds is 
established~\cite{Babu06,Babu07,Kovalchuk09a,Kovalchuk09b}. This produces a 
quasi-steady-state for $t \gtrsim 25000 $s, where the average cluster size 
equals 2.45 particles/cluster. It should be noted that the evolution of the 
small species slightly increases at long times. This effect is produced by 
gravity~\cite{Odriozola04} and is followed by an increase of the average cluster 
size (not captured) and a final depletion of the colloidal particles which 
accumulate at the flask bottom~\cite{Wu03,Agustin06a,Agustin06b}. 

When the particles are negatively charged, electrostatic forces are expected to 
hamper the specific accumulation of anions, which now act as co-ions, at the 
particle surface. For this reason, the average lifetime of primary bonds is 
expected to be less influenced by the anions nature. This is in agreement with 
the large and constant $\tau_1$ values shown in Table 1 for all electrolytes. 
Since we use sodium salts for all cases, the cation in solution is always the 
same independently of the salt employed, while the anions change. It hence 
follows that only counter-ions have an effect on $\tau_1$. From this result, it 
seems that anions are easily removed from the bonding area when the particles 
are negatively charged (note that this area is the co-ions less favorable 
electrostatic region to be placed). Conversely, $P_1$ and $\tau_2$ highly 
depends on the co-ion in solution, indicating that co-ions play an important 
role at slightly larger interparticle distances. The value of $\tau_2$ gradually 
decreases by increasing the dispersion contribution of the anions, suggesting 
that the secondary potential minimum is progressively shifted away from the 
particle surface, where the Hamaker force is smaller. This result emphasizes the 
important role of non-DLVO contributions on the PMF, even when the anions (the 
ions expected to specifically adsorb at the colloid surfaces) act as co-ions. 
Interestingly, $\tau_2$ attains smaller values in the anionic latex than in the 
cationic one. This could be due to the fact that a higher number of sodium ions 
are necessary to screen the more important effective charge of the anionic 
particles (the effective charge is expected to increase due to the specific 
anion adsorption). As a result, all secondary minima would shift away from the 
particle surface. Finally, $P_1$ is practically independent of the sign of the 
particles although strongly depend on the anion nature. 

In brief, following the series Cl$^-$, NO$_3^-$, SCN$^-$, the parameters 
obtained from the fitting procedure to the CSD suggest: i) An increasing trend 
for the PMF barrier for both, the cationic and the anionic system, according to 
the $P_1$ decreasing trend. ii) An increasing trend for the potential of mean 
force at contact, for the cationic system, and a practically invariant value for 
the anionic system. This is in correspondence with the obtained decrease of 
$\tau_1$ for the cationic system and the constant value of $\tau_1$ for the 
anionic colloidal particles. iii) A decreasing trend for the depth of the 
secondary minimum, in agreement with the decreasing values of $\tau_2$ for both 
latices. Additionally, the depth of the secondary minima for all electrolytes 
and for the anionic case should be smaller than those corresponding to the 
cationic case. iv) Finally, an increasing trend for the adsorption of anions for 
both systems. This is to agree with the increase of the mobility values obtained 
for the anionic system, as well as with the mobility reversal of the cationic 
particles (see the mobility data included in Table 1). The PMF from Monte Carlo 
simulations should capture these trends.

\section{Simulation details} \label{Sim}

\begin{figure}
\resizebox{0.5\textwidth}{!}{\includegraphics{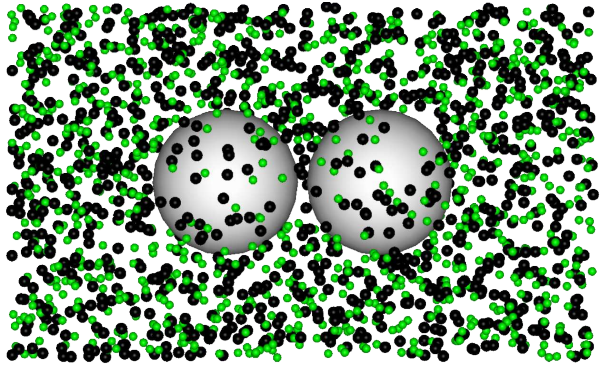}}
\caption{\label{snapshot} Snapshot of an equilibrated configuration of the 
system. Colloidal particles are white, light green particles are coions, and 
black particles are counterions.}
\end{figure}

Canonical Monte Carlo (MC) simulations are implemented for obtaining the PMF 
between two like-charged colloidal particles immersed in a 1-1 electrolyte. The 
macroparticles (colloidal particles) radius is $a_M=25\mbox{\AA}$ and carry the 
charge which corresponds to $\sigma = \pm50$ mC/m$^2$ at their center. Two 
colloidal particles are fixed in the simulation box at a surface-surface 
distance $h$ from one another. These particles remain fixed during a simulation 
run (only ions are allowed to explore the configuration space). Several 
surface-surface distances, $h$, are independently set as different runs to build 
the PMF. The simulation box is a prism having $L_x=L_y=120\mbox{\AA}$, and 
$L_z=250\mbox{\AA}$, sides much larger than the Debye-Hückel screening length 
for all studied cases. This condition is important to avoid size effects. The 
origin of coordinates is set at the prism center and periodic boundary 
conditions are set for the three directions. The 1-1 electrolyte is modeled by 
hard spheres of radius $a_c$ (cation) and $a_a$ (anion) with a centered point 
charge. As for the real experiment, an electrolyte concentration of $0.6$M is 
set. Additional ions are added to make the system electroneutral. Initially, 
these electrolyte particles are randomly placed avoiding overlaps. Similar 
system setups were employed elsewhere to study forces between fixed colloidal 
particles~\cite{Felipe05,Odriozola06}. The rout for obtaining the PMF is that 
described in references~\cite{Felipe05,Odriozola06,Tavares04,Bostrom06}. This 
type of simulations is frequently employed to compare the resulting PMF with 
those obtained by integral equations and density functional 
theories~\cite{Felipe05,Lima07,Jin11}. 

All excluded volume interactions are modeled by hard interactions. That is, 
overlaps are always rejected and non overlapping configurations are given a null 
excluded volume contribution to the configuration energy. On the other hand, the 
electrostatic contribution between any pair of sites $ij$, where $i$ and $j$ are 
either charged sites of the colloidal particles or ions, is given by
\begin{equation}
U_{el}=k_BTl_B \frac{z_i z_j}{r_{ij}} 
\end{equation}
where $r_{ij}$ is the distance between sites $i$ and $j$, and $z_i$ and $z_j$ 
are the valences of sites $i$ and $j$, respectively. The electrostatic strength 
is given by the Bjerrum length, i.~e., by 
\begin{equation}
l_B=\frac{e^2}{\varepsilon k_BT}
\end{equation}
where $\varepsilon$ is the dielectric constant. $l_B=7.14 \mbox{\AA}$ is set 
for water at $T=298$K. Finally, dispersion contributions are added to the 
configuration energy for the ion-macroparticle and for the ion-ion interaction. 
They read
\begin{equation}\label{Udisp}
\begin{array}{ll}
U^{im}_{disp}=-\frac{B_{im}}{r^3_{im}}\\
U^{ij}_{disp}=-\frac{B_{ij}}{r^6_{ij}}
\end{array} 
\end{equation}
being $B_{im}$ the dispersion parameter for the $i$-ion and the macroparticle 
(cation-macroparticle, anion-macroparticle) and $B_{ij}$ the dispersion 
parameter for the $ij$ ion-ion contribution (cation-cation, anion-anion, 
anion-cation). The values for these parameters are taken from Tavares et al.~and 
Bostr\"{o}m et al.~\cite{Tavares04,Bostrom06}. Electrostatic interactions are 
treated using the Ewald summation formalism. The convergence factor was fixed to 
5.6/$L_x$. There were set five reciprocal lattice vectors for $x$ and $y$ 
directions and six for the $z$ direction. A snapshot of an equilibrated 
configuration for positive colloidal particles and NaCl, (with $a_c=1.5 
\mbox{\AA}$, $a_a=2 \mbox{\AA}$) is shown in Fig.~\ref{snapshot}.

The effective electrostatic force acting on colloidal particle $m$ is obtained 
by simply accounting for all sites contributions, i.~e., by computing
\begin{equation}\label{electrostatic}
F_{el}=<\sum_{i} -\nabla U_{el}(r_{im}) >
\end{equation}
where $i$ runs over all sites except the corresponding colloidal particle site. 
The same procedure applies to the dispersion contribution, leading to 
\begin{equation}\label{dispersive}
F_{disp}=<\sum_{i} -\nabla U^{im}_{disp}(r_{im}) >.
\end{equation}

On the other hand, the contact force contribution (also called collision 
contribution~\cite{Tavares04,Bostrom06}) is obtained by integrating the ions 
contact density at the colloidal particle surface, $\rho_c$, i.~e., by means of
\begin{equation}\label{contact}
F_{c}=-\int_{S} \rho_c k_BT \hat{\mathbf{n}}ds
\end{equation}
In this case we approach $\rho_c$ at $ds$ by extrapolating the density of each 
species close to the surfaces. Finally, it should be mentioned that all these 
contributions to the net force are interdependent.

\section{Results}\label{Res}

\begin{figure}
\resizebox{0.5\textwidth}{!}{\includegraphics{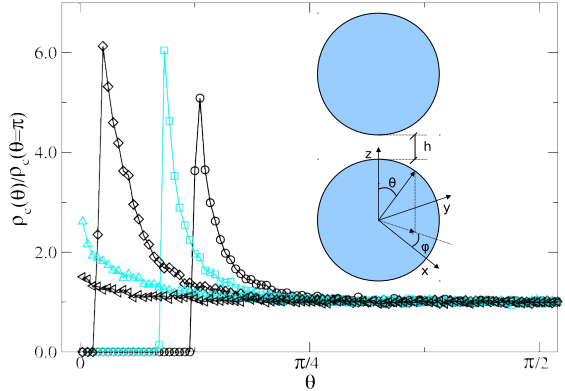}}
\caption{\label{gdrs-sup} Normalized counterion surface density, $\rho_c 
(\theta)/ \rho_c (\theta \!= \!\pi)$, as a function of the polar angle $\theta$ 
for the positive system under NaCl. Dark circles, light squares, dark diamonds, 
light triangles, and dark triangles correspond to $h=$0, 2, 4, 6, and 
10$\mbox{\AA}$, respectively. The inset shows the employed spherical coordinate 
system.}
\end{figure}

Simulations are performed to gain insight into the mechanisms through which 
different monovalent anions lead to different overall aggregation kinetics. 
Nonetheless, even without taking into account many details such as water 
structure, surface charge distribution, and roughness, among others, we can only 
study much smaller particles than the ones employed in the real experiments (100 
times smaller). Thus, only trends are expected to be comparable with the data 
obtained from experiments. 

As mentioned in section~\ref{Sim}, the PMF acting on both colloidal particles 
at a fixed distance can be accessed by ensemble averaging all its contributions. 
To understand their behavior it is convenient to first take a look at the ionic 
density profile on the surface of the colloidal particles. This profile is shown 
for the counterions (Cl$^-$) and for the cationic system in Fig.~\ref{gdrs-sup}. 
Different symbols correspond to different surface-surface separation distances, 
$h$. The inset of the same figure shows the definitions of angles $\theta$ and 
$\varphi$ (in spherical coordinates). By symmetry, surface ionic densities do 
not depend on $\varphi$. For $h=0$, there is an excluded region for 
$\theta\lesssim\pi/8$. That is, anions cannot enter in-between the colloidal 
particles. For slightly larger $\theta$, a large peak is produced, pointing out 
a strong counterion accumulation occurring at the surfaces of both colloidal 
particles (these peaks are absent for coions, Na$^+$, as expected). The reason 
for this to occur is twofold. On the one hand, in that region counterions are 
attracted by the electrostatic and dispersion forces of both particles 
(counterions are placed in contact to both macroparticles surfaces). In fact, 
this peak is placed where counterions minimize their electrostatic energy. On 
the other, the large surface/volume relationship of the region favors entropic 
adsorption (by increasing the accessible volume of other ions). For larger 
$\theta$ the counterion surface density monotonically decreases reaching a 
constant value for $\theta\gtrsim\pi/4$. The inhomogeneous distribution of ions 
around the dumbbell is responsible for the appearance of indirect forces between 
the macroparticles (see equations~\ref{electrostatic}-\ref{contact}). There is 
no net force acting on $x$ and $y$, as the ionic distribution is symmetric 
around the $z$ axis (independent of $\varphi$). The inhomogeneity in $\theta$ 
produces forces in the $z$ direction only. Hence, the large accumulation of 
counterions at both macroparticles surfaces should produce a large repulsive 
contact contribution to the overall interaction force, since these ions are 
pushing the colloidal surfaces away in order to enter the low-energy 
interparticle region, but, in turn, they should also attract the colloidal 
particles by electrostatic and dispersion forces (bridging). Conversely, the 
counterions adsorbed at large $\theta$ are producing contributions to the force 
in the opposite direction, and so, they may counterbalance the peak effect since 
they act on a larger surface area (there is no excluded region at large 
$\theta$). The net force is the sum of these intricate contributions to the 
direct colloid-colloid interaction. 

As the macroparticles separation distance $h$ increases, the surface density 
peak grows and shifts to smaller values of $\theta$. These two facts would yield 
larger contributions of the forces in the $z$ direction. For $h=2a_a$ (the 
counterion diameter), the height of the peak reaches a maximum, decreasing for 
larger values of $h$. The peak is now placed in the inter-particle region, i. 
e., at the point of zero electric field (center of the simulation box) where 
counterions minimize their electrostatic energy. Thus, for $h=2a_a$, the 
counterions can access all macroparticles surfaces and the in-between excluded 
region disappears. For $h>2a_a$ the peak's height rapidly decreases as $h$ 
increases. However, it completely vanishes at large $h$ where the double layers 
become totally independent of each other and the net colloidal forces fade out. 

\begin{figure}
\resizebox{0.5\textwidth}{!}{\includegraphics{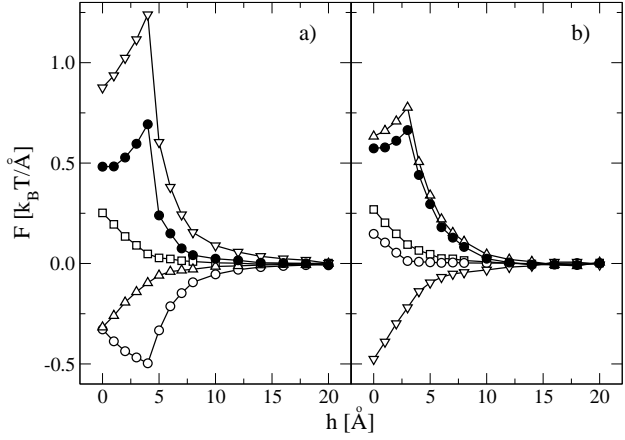}}
\caption{\label{FNaCl} Forces acting on the colloidal particles as a function 
of their surface-surface separation distance, $h$. a) For positive colloidal 
particles. b) For negative colloidal particles. Squares, circles, triangles up, 
and triangles down correspond to the electrostatic, dispersive, Na$^+$-contact, 
and Cl$^-$-contact, force contributions. Bullets correspond to the net force. }
\end{figure}

For NaCl with $a_c=1.5\mbox{\AA}$ and $a_a=2.0\mbox{\AA}$, the force 
contributions are shown in Fig.~\ref{FNaCl} a) for cationic colloidal particles, 
and in Fig.~\ref{FNaCl} b) for anionic colloidal particles. Let's focus first in 
Fig.~\ref{FNaCl} a). A positive (repulsive) electrostatic contribution for all 
distances is seen. This contribution is monotonously decreasing and reaches 
values close to zero for $h \simeq 10\mbox{\AA}$. In other words, the direct 
macroparticle-macroparticle electrostatic contribution is fully screened for 
distances larger than a few ion diameters. For the given conditions, i.~e., for 
large electrolyte concentrations (0.6 M), this contribution is the smallest. The 
dispersion contribution to the net force is mostly related to the 
anion-colloidal particle interaction, since cations have a small dispersion 
parameter (see equation~\ref{Udisp}) and they poorly adsorb onto the positive 
colloidal particle surface. This contribution is always negative (attractive), 
and, as explained in the previous paragraphs, is related to the large anion 
concentration located in-between the colloidal surfaces (as shown in 
Fig.~\ref{gdrs-sup}). When the peak of the counterion surface density profile is 
at its maximum, i.~e., at $h=2a_a$, the dispersion force yields a minimum. This 
points out that those anions in-between the particles attract them towards the 
simulation box center, producing the effect of a bridge. However, the contact 
contribution produced by this high local anion concentration has exactly the 
opposite behavior. That is, it yields a positive contact contribution which also 
peaks at $h=2a_a$. This contribution is larger than the bridging effect caused 
by the dispersion force. Finally, the contact cation contribution is positive 
since cations preferably locate at the outside of the interparticle region. For 
a large enough $h$ the ionic surface distributions become homogeneous and all 
contributions disappear. The sum of all contributions to the force is seen in 
Fig.~\ref{FNaCl} as bullets, which turns out to be repulsive, peaking at 
$h=2a_a$. As can be seen, the dominant contribution is the contact repulsive 
force that counterions exert on the macroparticle surface. All contributions 
are, however, interdependent. 

Fig.~\ref{FNaCl} b) shows the data obtained for the anionic colloidal particles 
at the same electrolyte conditions. The electrostatic contribution to the net 
force is very similar to the cationic case. That is, the contribution is always 
repulsive, shows a monotonously decreasing behavior, is fully screened for 
distances larger than two ion diameters, and, in general, shows similar values 
than the cationic system. Conversely, the dispersion component behaves very 
differently than for the positive macroparticles. This component is repulsive 
and monotonously decreasing for the anionic case, contrasting the attractive 
dispersion contribution obtained for the positive system. This is due to the 
fact that anions, which produce the largest dispersion contribution, are now far 
from the interparticle region, and are mostly adsorbed on the external surface 
of both particles. Hence, they pull the particles away from each other, yielding 
a repulsive contribution. This is the most important difference between both 
cases. In fact, the counterions contact contribution is positive and the coions 
contribution is negative, both showing similar trends than for the cationic 
system. However, for the anionic system the counterion contact repulsion is 
smaller and the coion contact attraction larger. This is due to the smaller 
adsorption of Na$^+$ than Cl$^-$, which in turn is explained by the larger 
dispersion parameter of chloride. These differences in the strength of the 
contact contributions counterbalance the sign change of the dispersion component 
in such a way that the net force of both systems turns out to be very similar. 
This does occur for $a_c= 1.5\mbox{\AA}$, and $a_a=2\mbox{\AA}$, but it is not 
general. In fact, we tuned $a_c$ for a fixed $a_a$ to obtain similar potentials 
of mean force. This was done since the experimental overall aggregation rate is 
practically equal for both systems (anionic and cationic) under 0.6 M of NaCl, 
and so, similar potentials of mean force are expected. Probably, larger $a_c$ 
and $a_a$ values also yield similar potentials of mean force (note that $a_c$ is 
smaller than the generally accepted value $a_c \approx 2\mbox{\AA}$). 

\begin{figure}
\resizebox{0.5\textwidth}{!}{\includegraphics{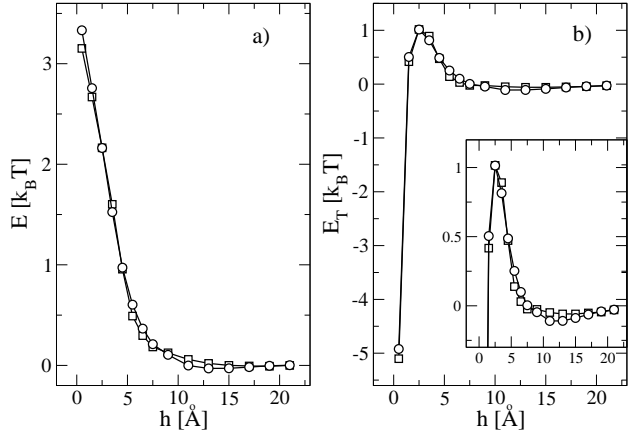}}
\caption{\label{ENaCl} a) Potential of mean force, $E(h)$, as obtained by 
integration of the total force as a function of $h$. b) The same data plus the 
macroparticle-macroparticle Hamaker contribution of polystyrene in water, 
$E_T(h)$. The inset zooms in the data of panel b). }
\end{figure}

Forces are easily translated into potentials of mean force by means of
\begin{equation}
E(h)=\int_{\infty}^{h}F_t(x)dx.
\end{equation}
$E(h)$ is plotted in Fig.~\ref{ENaCl} a) as a function of the separation 
distance $h$ for both systems. As mentioned, $E(h)$ is  similar for the cationic 
and anionic systems for all $h$. In Fig.~\ref{ENaCl} b) it is plotted the same 
data plus the Hamaker contribution, $E_T(h)=E(h)+E_H(h)$, which 
reads~\cite{Hunter} 
\begin{equation}\label{E_H}
\!E_H(h)\!=\!\frac{-A_H}{6} \!\! \left[\!\frac{2a_M^2}{h(4a_M\!+\!h)}\!+ \!\frac{2a_M^2}{(2a_M\!+\!h)^2}\!+\! \ln \!\left(\!\frac{h(4a_M\!+\!h)}{(2a_M\!+\!h)^2}\! \right) \right]
\end{equation}
being $A_H=0.95\times10^{-20}$J the Hamaker constant for polystyrene in water 
and $a_M=25\mbox{\AA}$ the colloidal particle (macroparticle) radius. Panel b) 
of Fig.~\ref{ENaCl} shows the existence of a potential barrier peaking at 
$h=2.5\mbox{\AA}$ even for a 0.6 M electrolyte concentration. This contrasts the 
DLVO theory predictions (no barrier for this salt concentration). As was pointed 
out, the potential barrier is related to the accumulation of counterions around 
the surface-surface contact region. That is, work must be done by or on the 
system in order to release the counterions from the very low energy region 
in-between the particles surfaces to relocate them in a less favorable place. 
According to the simulation data this work is not compensated by the gain of the 
Hamaker contribution. This result agrees with the experimental values found for 
the probability of forming a primary bond, $P_1$, which are for all cases 
smaller than one. If this were true, the dimmer formation rate constant $k_{11}$ 
would approach better the theoretical Smoluchowsky value, $k^{Smol}_{11} = 11.1 
\times 10^{-18}$ m$^3$/s (water at 293 K), and the overall effect of 
hydrodynamic interactions would be less important than generally accepted 
(hydrodynamic interactions are said to decrease the Smoluchowsky value in a 
factor of two~\cite{spielman70,honig71}).    

It should also be noted that the secondary minima shown in Fig.~\ref{ENaCl} b) 
are not deep enough to produce relatively stable secondary bonds. This is 
expected for small particles as the ones considered for the simulations. For 
much larger particles, as these employed to obtain the experimental data shown 
in Fig.~\ref{CSD}, the Hamaker contribution enlarges producing the well known 
secondary minimum. Additionally, according to the fitted $\tau_2$ parameter, the 
secondary minimum for the cationic case should be deeper than the one 
corresponding to the anionic particles. This is not captured by the simulations. 
Finally, the obtained primary minima are too deep to allow for bond breakup. 
Note that equation~\ref{E_H} diverges for $h\rightarrow0$ and thus it surely 
overestimates the real Hamaker contribution for very short distances (first 
point of panel b) of Fig.~\ref{ENaCl} is evaluated at $h=0.1 \mbox{\AA}$). 
Additionally, other contributions are expected to be relevant at these very 
short distances (for instance, water molecules hydrating surface charges must 
also be released from the in-between region to produce a bare-bare bond).   

%Nevertheless, the depth of the primary minimum is still too large to justify a reversible process. Thus, particle roughness, solvent structuring, partial dehydration of adsorbed ions, and local surface charges (instead of the central point charge and the very different size of the colloidal particles employed in the simulations) may be responsible of a shallower minimum. As was already mentioned, only trends are being analyzed.

\begin{figure}
\resizebox{0.5\textwidth}{!}{\includegraphics{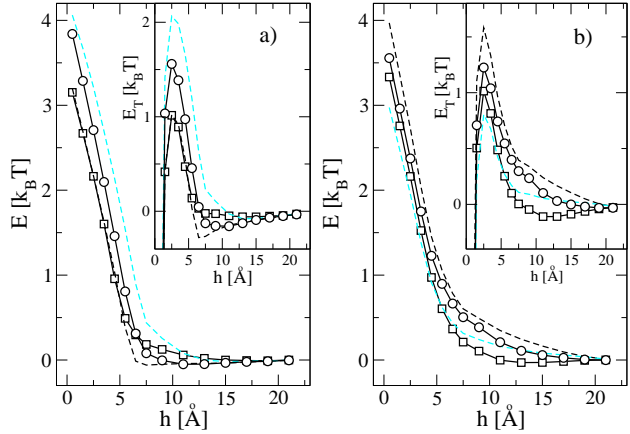}}
\caption{\label{ENaSCN} Potential of mean force, $E(h)$, as a function of the 
separation distance, $h$, for a) the cationic system, and b) the anionic system. 
The insets show the same data plus the Hamaker contribution, $E_T(h)$. Squares 
correspond to NaCl, whereas circles correspond to NaSCN with 
$a_a=2.75\mbox{\AA}$. Dark dashed and light dashed lines correspond to NaSCN 
with $a_a=2.50\mbox{\AA}$ and $3.00\mbox{\AA}$, respectively.}
\end{figure}

Up to this point the calculations only considered NaCl as the dissolved salt. 
From here on we focus on the results for NaSCN for different SCN$^-$ hydrated 
radius, $a_a$, keeping constant the fitted Na$^+$ radius (NO$_3^-$ is expected 
to show an intermediate behavior between Cl$^-$ and SCN$^-$ and so, it is not 
considered for the simulations). As mentioned at the end of section~\ref{PBF},  
the parameters obtained from the fitting procedure to the aggregation data 
suggest the following changes of $E_T(h)$ when comparing the NaCl with the NaSCN 
cases: i) An increase of the $E_T(h)$ peak for both, the cationic and the 
anionic system, in correspondence with the $P_1$ decrease when aggregation is 
induced by NaSCN. This increase should be similar for both systems ($P_1$ 
decreases similarly in both systems). ii) An increase of the potential contact 
value, $E_T(h=0)$, for the cationic system, and a similar value of $E_T(h=0)$ 
for the anionic system. This is in correspondence with the obtained decrease of 
$\tau_1$ for the cationic system and the constant value of $\tau_1$ for the 
anionic colloidal particles. iii) Next, the smaller values of $\tau_2$ found for 
NaSCN would translate in a decrease of the depth of the secondary minimum for 
NaSCN. This would also apply for both latices. In addition, the depth of the 
secondary minima for all electrolytes and for the anionic case should be smaller 
than those corresponding to the cationic case. iv) Finally, the simulations 
should also produce a greater adsorption of the SCN$^-$ ion for both systems. 
This is to agree with the increase of the mobility values obtained for the 
anionic system when changing from NaCl to NaSCN electrolyte, as well as with the 
mobility reversal of the cationic particles (see the mobility data included in 
Table 1).   

The results of the calculations involving the NaSCN are given in 
Fig.~\ref{ENaSCN}. For an easy comparison, the data obtained for the NaCl are 
also included as squares. Fig.~\ref{ENaSCN} a) corresponds to the cationic 
system and Fig.~\ref{ENaSCN} b) to the anionic one. The insets show the same 
energy data plus the Hamaker contribution. From Fig.~\ref{ENaSCN} a) it is seen 
that the anionic radius, $a_a$, must be larger than $2.5\mbox{\AA}$ to obtain a 
higher repulsive barrier and a higher $E_T(h=0)$ for NaSCN than for NaCl. This 
is so since the PMF of the positive colloidal particles increases with the anion 
size. On the contrary, Fig.~\ref{ENaSCN} b) shows that the PMF of the anionic 
system decreases with the SCN$^-$ size, producing a smaller energetic barrier 
than the NaCl reference for $a_a<2.5\mbox{\AA}$. Thus, according to the model, 
SCN$^-$ should have a hydrated size ranging in [2.5; 3.0]$\mbox{\AA}$ to match 
the $P_1$ decrease found for both latices from the master equation fits. This is 
a reasonable range for the hydrated size of SCN$^-$. Fig.~\ref{ENaSCN} includes 
the calculations for $a_a=2.75\mbox{\AA}$ (open circles). For this anionic size 
both panels of Fig.~\ref{ENaSCN} show energetic barriers larger than those 
obtained with NaCl (insets of Fig.~\ref{ENaSCN}), in agreement with point i) of 
previous paragraph. Additionally, for the cationic system (panel a)) $E_T(h=0)$ 
is clearly larger for SCN$^-$ than for Cl$^-$, whereas the increase is less 
pronounced for the anionic system (panel b)). So, point ii) of previous 
paragraph is qualitatively matched. Point iii) is partially matched. That is, 
for the cationic system, the depth of the secondary minima decreases only for 
NaSCN with $a_a=3.00\mbox{\AA}$, but not for $a_a=2.75\mbox{\AA}$ as it should. 
On the other hand, the secondary minima for the cationic system are deeper than 
for the anionic system for NaSCN, which is right. In fact, the secondary minimum 
disappears for the anionic system and the broadness of the energetic barrier 
turns significantly larger. This longer range of the PMF barrier suggests that 
pairs of counter and coions must be released from the in-between surface-surface 
region to produce a stable bond. Summarizing, in general and up to this point, 
the qualitative agreement between the suggested trends from the fitted 
parameters and the simulation results is good. This enhances confidence in both 
treatments.       

\begin{figure}
\resizebox{0.5\textwidth}{!}{\includegraphics{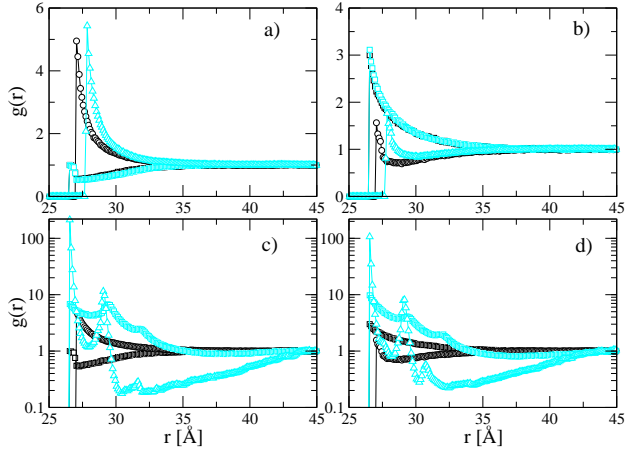}}
\caption{\label{gdrs} Radial distribution functions, $g(r)$, for Na$^+$ 
(squares), Cl$^-$ (circles), and SCN$^-$ ions (triangles), with respect to an 
isolated positive a) and negative b) macroparticle. Black lines and symbols 
correspond to NaCl runs and light (cyan) lines and symbols correspond to NaSCN 
runs with $a_a=2.75\mbox{\AA}$. Panels c) and d) show the data corresponding to 
a SCN$^-$ hydrated radius of $a_a=1.5\mbox{\AA}$. As reference, these panels 
also include the NaCl data shown in panels a) and b). }
\end{figure}

Unfortunately, point iv) is not fulfilled. The radial distribution functions 
obtained for an isolated colloidal particle immersed in NaCl electrolyte 
solution and in a NaSCN electrolyte solution (with $a_a=2.75\mbox{\AA}$) show no 
practical differences for the adsorption of SCN$^-$ and Cl$^-$. This is shown in 
panels a) and b) of Fig.~\ref{gdrs} for the positive and negative systems, 
respectively. This suggests that some SCN$^-$ anions should be partially losing 
their water shells in order to attach the colloidal particles surfaces. To 
confirm this solvent mediated mechanism simulations explicitly accounting for 
the solvent molecules are needed (recently, potentials of mean force were build 
up directly from force fields~\cite{Kalcher10,Fyta10}). Nonetheless, with the 
employed model we can explore the effect of the SCN$^-$ hydrated size on their 
adsorption on the colloidal particles surfaces. For this purpose an extra 
calculation is made for an isolated colloidal particle immersed in SCN$^-$ with 
$a_a=1.5\mbox{\AA}$. This would represent the size of a partially hydrated 
SCN$^-$ ion. Results are shown in panels c) and d) of Fig.~\ref{gdrs}. These 
panels show a very large adsorption of SCN$^-$ for both cases (positive, a), and 
negative, b), colloidal particles). Furthermore, both, counterions and coions 
radial distribution profiles show several peaks which reveals the appearance of 
charge reversal~\cite{Felipe08,Martin-Molina09a} (for the positive chase) and 
overcharging~\cite{Mesina09,Boinovich10} (for the negative case) phenomena. 
Indeed, the contact peak of the radial distribution function for the positive 
system is produced by the adsorption of approximately 90 anions. For the anionic 
system, the number of absorbed SCN$^-$ ions average 55. This leads to an 
effective surface charge density at $3\mbox{\AA}$ from the surface close to -100 
mC/m$^2$ (accounting for both, the adsorbed anions and cations) for the cationic 
system and -108 mC/m$^2$ for the anionic one. That is, the effective surface 
charge density of the anionic system double (overcharging), and the cationic 
system not only change sign (charge reversal), but also double its original 
absolute value. Thus, the adsorption is strongly overestimated by these 
calculations which signal the extreme sensitivity to the considered hydrated 
radius of the ions~\cite{Martin-Molina09a,Martin-Molina09b,Quesada-Perez10} 
(sensitivity to this parameter is strongly enhanced when including the 
dispersion contribution). However, since the dehydrating process is 
energetically demanding, not all the SCN$^-$ ions placed close to the colloidal 
particle surface are expected to follow this rout. Consequently, a significant 
but not very large amount of ions should dehydrate while adsorbing, explaining 
the mobility measurements, whereas at larger separations from the colloidal 
particles surfaces, anions would be fully hydrated to produce forces such as 
those obtained for a SCN$^-$ radius of $2.75\mbox{\AA}$. These adsorbed and 
partially dehydrated ions should also increase the potential energy at very 
short distances, which aids explaining the full reversibility of the primary 
bonds for the cationic system. Probably these ions are not being totally removed 
from the surfaces while forming a primary bond leading to their occlusion. This 
phenomenon was recently proposed (for hydronium ions) to explain the observed 
reduction of surface charges during the aggregation and coalescence of elastomer 
particles~\cite{Gauer10}.

\section{Conclusions} \label{Conc}

Population balance fitting of experimental aggregation data and potentials of 
mean force from simulations support the existence of an energetic barrier for 
the potential of mean force between hydrophobic colloidal particles at high 
electrolyte concentrations. This is found not only for NaSCN but even for NaCl, 
although the barrier is smaller in this last case. Furthermore, positive and 
negative colloids show the same increasing trend for the height of the energetic 
barrier following the series NaCl, NaNO$_3$, NaSCN. These findings contrast the 
DLVO predictions. For positive particles, the energetic barrier would be 
produced by the work needed for releasing the adsorbed counterions from the 
in-between surface-surface region and relocating them in a not so energetically 
favorable place. Thus, the barrier would be located at very short 
surface-surface distances. In the case of negative colloids, the barrier extends 
to larger distances suggesting that pairs of counter and coions must be released 
from the in-between surface-surface region to produce a stable bond. According 
to simulations and population balance fitting, ions like SCN$^-$, which show a 
natural tendency to adsorb onto hydrophobic surfaces, produce a larger energetic 
barrier for positive and negative surfaces. In the case of the positive 
colloidal particles, SCN$^-$ produces weaker primary bonds yielding a clear 
reversibility of the aggregation processes.  

\section{Acknowledgements}
The author thanks fruitful and enrichment discussions with Drs.~Teresa 
L\'{o}pez-Le\'{o}n, Juan Manuel L\'{o}pez-L\'{o}pez, Delfi Bastos-Gonz\'{a}lez, 
Juan Luis Ortega-Vinuesa, and Manuel Quesada-Perez.

%\bibliography{SPOS}

%

\end{document}